\documentclass[11pt]{article}
\usepackage[dvips]{graphicx}
\begin{document}
\title{Domain-size control by global feedback in bistable systems}
\author{Hidetsugu  Sakaguchi\\
Department of Applied Science for Electronics and Materials,\\ Interdisciplinary Graduate School of Engineering Sciences,\\
Kyushu University, Kasuga, Fukuoka 816-8580, Japan}
\maketitle
\begin{abstract}
We study domain structures in bistable systems such as the Ginzburg-Landau 
equation. The size of domains can be controlled  by a global negative feedback.  
The domain-size control is applied for a localized spiral pattern.
\end{abstract}
\clearpage
Spatially localized states have been observed experimentally
in binary fluid mixtures \cite{rf:1},
electroconvection in nematic liquid crystals \cite{rf:2} and in granular media undergoing the Faraday instability \cite{rf:3}.
Some simple model equations have been studied to understand the mechanism
of the localized states found in dissipative systems. 
Soliton-like localized states have been found in the quintic 
complex Ginzburg-Landau equation and the coupled complex Ginzburg-Landau equations \cite{rf:4,rf:5}. Wormlike localized states were studied with the 
anisotropic complex Ginzburg-Landau equation coupled with a scalar mode \cite{rf:6}. A self-trapping mechanism works for localized states in the quintic Swift-Hohenberg equation \cite{rf:7}. 
The long-range inhibition is important for localized states in some reaction diffusion equations \cite{rf:8}.
On the other hand, controlling  chaotic dynamics has been investigated 
with the OGY method and the feedback method \cite{rf:9,rf:10}. 
Zykov et al. studied the control of spiral waves in a spatially extended system  by a global feedback \cite{rf:11}.  

We study the control of the domain size of localized domains 
in spatially extended bistable systems. 
Our first model equation is based on the Ginzburg-Landau equation coupled with 
an inhibitory medium. 
\begin{eqnarray}
\frac{\partial u}{\partial t}& =&u-u^3-v+\frac{\partial^2 u}{\partial x^2},\nonumber\\
\tau \frac{\partial v}{\partial t}& =&-v+K(u-c)+D\frac{\partial^2 v}{\partial x^2},
\end{eqnarray}
where $u$ obeys the Ginzburg-Landau equation, $v$ denotes the inhibitory variable, $D$ is the diffusion constant of $v$, $\tau$ is the time constant of $v$ and $c$ is a parameter between -1 and 1. This model equation is a reaction diffusion equation. The system size is $L$ and the periodic boundary condition or the no-flux boundary condition is assumed.
If $D$ is sufficiently large, a localized state is obtained as in [8].
If $D$ is infinitely large, $v$ is uniform for $0<x<L$, that is, $v=\langle v\rangle=(1/L)\int_0^Lvdx$. The second equation in Eq.~(1) is reduced to 
\begin{equation}
\tau \frac{d\langle v\rangle}{dt}=-\langle v\rangle+K(\langle u\rangle-c),
\end{equation}
where $\langle u\rangle=(1/L)\int_0^L udx$.
If the adiabatic approximation is assumed, $\langle v\rangle=K(\langle u\rangle-c)$. 
Substitution of this relation into the first equation of Eq.~(1) yields
\begin{equation}
\frac{\partial u}{\partial t}=u-u^3-K(\langle u\rangle-c)+\frac{\partial^2 u}{\partial x^2},
\end{equation}
where the third term on the right-hand side represents the global negative feedback.  If the third term $-K(\langle u\rangle-c)$ takes a small constant value $b$, there are two stable states near $u\sim\pm 1$. If the initial condition takes a domain structure, the domain wall moves 
with a constant velocity for nonzero $b$. 
The domains of positive (or negative) $u$ become dominant if $b$ is  positive (or negative).
In our negative feedback model with $K>0$, the effective control parameter $-K(\langle u\rangle-c)$ 
decreases (increases) as the domain size of positive (negative) $u$  increases.  Finally, $\langle u\rangle=c$ is attained and the domain growth stops. 
Then, the size of the domain of $u=\pm 1$ is approximately $(1\pm c)L/2$.
Namely the parameter $c$  determines the domain size.  We can control the domain size by changing the parameter $c$.  Figures 1(a) and (b) display the time evolution of the domain structure at $K=0.5,c=-0.2$ and $L=200$. The initial condition is $u(x)=1$ for $98<x<102$ and $u=-1$ for $x<98$ and $x>102$. 
The numerical simulation was performed with the pseudospectral method of mode-number 1024 and timestep 0.01.
\begin{figure}[htb]
\begin{center}
\includegraphics[width=6cm]{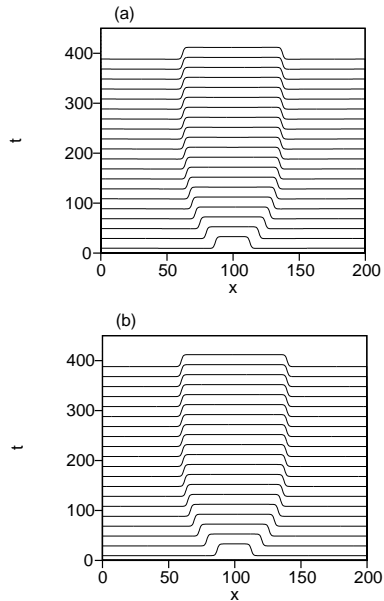}
\caption{(a) Time evolution of $u(x,t)$ by Eq.~(1) at $\tau=1,\;K=0.5,\;c=-0.2,\;D=10000$ and $L=200$. 
(b) Time evolution of $u(x,t)$ by Eq.~(3) at $K=0.5,\;c=-0.2$ and $L=200$. 
} 
\label{fig:1} 
\end{center}
\end{figure} 
Figure 1(a) displays the time evolution for Eq.~(1) with $D=10000$ and $\tau=1$.  A localized state with size 75.7 is obtained.  The domain size is calculated as a width of the region of $u>0$. Figure 1(b) displays the time evolution by the Ginzburg-Landau equation (3) with the global negative feedback.  A domain with a fixed size is obtained as a stationary state. 
The final size of the domain $u=1$ is 80.  Since the diffusion constant $D=10000$ in Eq. (1) is very large, the time evolution by Eq.~(1) is close to that by Eq.~(3). 
The domain size is well approximated at the value $(1+c)L/2=80$.

We can control the domain size even if the system has three stable states. 
The model equation is the quintic Ginzburg-Landau equation:
\begin{equation}
\frac{\partial u}{\partial t}=-au+u^3-u^5+e+\frac{\partial^2 u}{\partial x^2},
\end{equation}
where $a$ and $e$ are parameters. If $e=0$ and $a=3/16$, the potential energy 
$U=au^2/2-u^4/4+u^6/6-eu$ takes the same local minimum value 0 at $u=0$ and $\pm \sqrt{3}/2$. Two kinds of domain walls which connect $-\sqrt{3}/2$ and 0, and 0 and $\sqrt{3}/2$  do not move at the parameters $e=0$ and $a=3/16$.  For the other parameter values, the domain walls have finite velocities. 
We assume a model with a global negative feedback as
\begin{equation}
\frac{\partial u}{\partial t}=u^3-u^5-K_1(\langle u\rangle-c_1)-K_2(\langle u^2\rangle-c_2)u+\frac{\partial^2 u}{\partial x^2},
\end{equation}
where $\langle u\rangle=(1/L)\int_0^Ludx,\langle u^2\rangle=(1/L)\int_0^Lu^2dx$, and $K_1,K_2$ are 
coupling constants and $c_1,c_2$ are the control parameters which determine the domain sizes. 
Since the parameter value $(a,e)=(0,0)$ in Eq.~(4) is a codimension 2 point,  
we need two types of negative feedback.
If the negative feedback succeeds, the final state satisfies 
$\langle u\rangle=c_1$ and $K_2(\langle u^2\rangle-c_2)=3/16$. 
We have performed a numerical simulation for $L=200$  under the no-flux boundary condition. The initial condition is $u(x)=-1+2x/L$.
\begin{figure}[htb]
\begin{center}
\includegraphics[width=6cm]{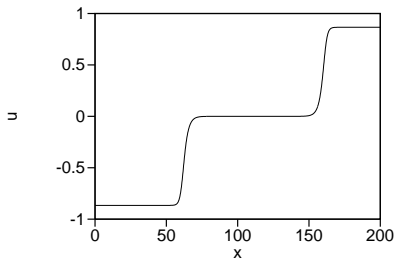}
\caption{Final stationary state for Eq.~(5) at $K_1=K_2=0.5,\;c_1=-0.1,\;c_2=0$ and $L=200$. } 
\label{fig:2} 
\end{center}
\end{figure} 
Figure 2 displays the final state in the time evolution by Eq.~(5) at $K_1=0.5,K_2=0.5,c_1=-0.1$ and $c_2=0$. There appear three domains of $u\sim -\sqrt{3}/2,\;0$ and $\sqrt{3}/2$.  
If the domain sizes are denoted as $l_+,l_0$ and $l_-$ respectively for the three domains of $u=\sqrt{3}/2,0$ and $-\sqrt{3}/2$, the domain sizes satisfy approximately  
$\langle u\rangle=\sqrt{3}/2(l_+-l_-)/L=c_1=-0.1$ and $\langle u^2\rangle=3/4(l_++l_-)/L=3/(16K_2)+c_2=3/8$.  The expected domain sizes are $l_+=38.4,l_0=100$ and $l_-=61.6$ for the parmeters $c_1=-0.1,c_2=0$ and $L=200$. 
The numerical result is approximately $l_+\sim39.7,l_0\sim 97.5$ and $l_-\sim 62.7$. The global feedback succeeds in the domain size control for this quintic Ginzburg-Landau system.

The same method is applicable for nonvariational systems.
We use the quintic complex Ginzburg-Landau equation:
\begin{equation}
\frac{\partial W}{\partial t}=-aW+(1+ic_2)\mid W\mid^2 W-\mid W\mid^4 W+\frac{\partial^2 W}{\partial x^2},
\end{equation}
where $W$ is a complex variable and $c_2$ is a nonvariational parameter. 
There are two stable uniform states: $W=0$, and $W=W_0\exp(i\omega_0 t)$ where $W_0=\sqrt{(1+\sqrt{1-4a})/2}$ and $\omega_0=c_2(1+\sqrt{1-4a})/2$. 
There exists a domain wall which connects the zero state and the oscillating state. 
\begin{figure}[htb]
\begin{center}
\includegraphics[width=6cm]{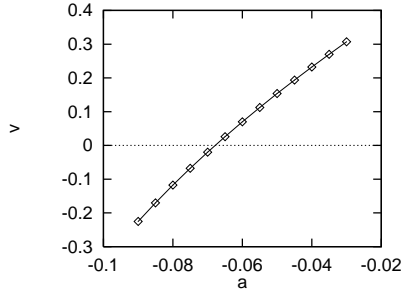}
\caption{Velocity $v$ of the domain wall as a function of $a$ for Eq.~(6) at $c_2=0.4$.} 
\label{fig:3} 
\end{center}
\end{figure} 
Figure 3 displays the numerically obtained velocity of the domain wall as a function of $a$ for $c_2=0.4$. The positive velocity implies that the oscillating state invades the zero state.  The velocity of the domain wall takes 0 at $a=a_c\sim 0.0678$.  The model equation with a global feedback is 
\begin{equation}
\frac{\partial W}{\partial t}=-aW+(1+ic_2)\mid W\mid^2 W-\mid W\mid^4 W-K\langle\mid W\mid^2\rangle W+\frac{\partial^2 W}{\partial x^2},
\end{equation}
where $K$ denotes the feedback strength and $\langle \mid W\mid^2\rangle=(1/L)\int_0^L\mid W\mid^2dx$. Figure 4 displays the time evolution of $\mid W\mid$ for $a=0.01,L=400,c_2=0.4$ and $K=0.12$. The initial condition is Re$W(x)=1$ and Im$W(x)=0$ for $196<x<204$, and $W(x)=0$ for $x<196$ and $x>204$.  The domain size $l$ of the oscillating state is calculated by the relation $a+K\langle\mid W\mid^2\rangle\sim a+K\mid W_1\mid^2l/L\sim a_c$, where $\mid W_1\mid\sim 0.866$ is the amplitude  of the oscillating state coexisting with the zero state at $a=a_c$ for Eq.~(6). The estimated value is $l=L(a_c-a)/(K\mid W_1\mid^2)\sim 256.8$. The numerically obtained size is approximately 257.  We can control the domain size by changing the parameter $a$ or $K$. 
\begin{figure}[htb]
\begin{center}
\includegraphics[width=6cm]{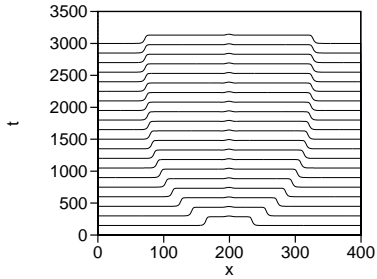}
\caption{Time evolution of $\mid W\mid$ for Eq.~(7) at $a=0.01,\;c_2=0.4,\;K=0.12$ and $L=400$.} 
\label{fig:4} 
\end{center}
\end{figure}

We have performed a simulation of the two-dimensional quintic complex Ginzburg-Landau equation with the global feedback.
\begin{equation}
\frac{\partial W}{\partial t}=-aW+(1+ic_2)\mid W\mid^2 W-\mid W\mid^4 W-K\langle\mid W\mid^2\rangle W+\nabla^2 W,
\end{equation}
where $K$ denotes the feedback strength and $\langle \mid W\mid^2\rangle=(1/L^2)\int_0^L\int_0^L\mid W\mid^2dxdy$. The parameters are $L=200,a=-0.05,c_2=0.4$ and $K=0.4$. The numerical simulation was performed with the psudospectral method of mode-number $256\times 256$ and timestep 0.005. The initial condition was Re$W(x,y)=0.0033(x-L/2)(r_d-r)$ and Im$W(x,y)=0.0033(y-L/2)(r_d-r)$ for $r<r_d$ where $r=\sqrt{(x-L/2)^2+(y-L/2)^2}$ and $r_d=55$, and $W=0$ for $r>r_d$. That is, a topological defect is set at the center $(x,y)=(L/2,L/2)$ as an initial condition. A spiral pattern evolves from the initial condition. The spiral pattern occupies only a finite domain by the global feedback. 
\begin{figure}[htb]
\begin{center}
\includegraphics[width=8cm]{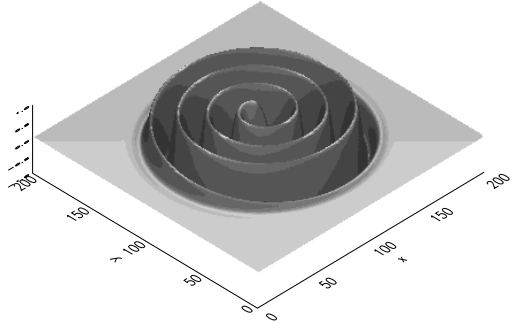}
\caption{3D-plot of Re$W$ for a localized spiral pattern as a result of  the time evolution by  Eq.~(8) at $a=-0.05,\;c_2=0.4,\;K=0.4$ and $L=200$.
} 
\label{fig:5} 
\end{center}
\end{figure} 
Figure 5 displays a 3D-plot of Re$W(x,y)$ for the localized spiral pattern. 
The value of $a+K\langle \mid W\mid^2\rangle$ is approximately -0.0648 for the stationary localized spiral, and the value is close to $a_c$, but it is slightly larger than $a_c$. It is probably due to the surface tension effect.  
Recently a localized spiral pattern is studied with the quintic complex Ginzburg-Landau without the global feedback term \cite{rf:12}. However, the localization mechanism is different.  In our model, the domain size of the spiral pattern can be controlled by the parameter $a$ or $K$.

To summarize, we have studied the control of the domain size for the 
Ginzburg-Landau type equations with a global negative feedback. 
The domain-size control by the global negative feedback is one of the simplest examples of the control for spatially extended dynamical system.    
The domain size is determined by the relation that the 
velocity of the domain wall is zero. 
The global negative feedback can be derived from the coupled reaction diffusion equation. In the crystal growth problem, the temperature field plays a role of the inhibitory medium through the latent heat released at the solidification. The global negative feedback  appears more naturally in electric circuits \cite{rf:13}. 
The control of the domain size is fairly robust and may be applicable for 
many systems.

\end{document}